\documentclass[runningheads]{llncs}
\usepackage{booktabs} 
\usepackage{lmodern}

\usepackage{longtable}
\usepackage{footmisc}
\usepackage{graphicx}
\usepackage{xcolor}

\usepackage{hyperref}
\expandafter\def\expandafter\UrlBreaks\expandafter{\UrlBreaks
  \do\a\do\b\do\c\do\d\do\e\do\f\do\g\do\h\do\i\do\j%
  \do\k\do\l\do\m\do\n\do\o\do\p\do\q\do\r\do\s\do\t%
  \do\u\do\v\do\w\do\x\do\y\do\z\do\A\do\B\do\C\do\D%
  \do\E\do\F\do\G\do\H\do\I\do\J\do\K\do\L\do\M\do\N%
  \do\O\do\P\do\Q\do\R\do\S\do\T\do\U\do\V\do\W\do\X%
  \do\Y\do\Z}

\begin{document}
\title{Bounded Temporal Fairness\\ for FIFO Financial Markets}

\author{Vasilios Mavroudis}
\authorrunning{V. Mavroudis}
\institute{University College London\\
\email{v.mavroudis@cs.ucl.ac.uk}}

\maketitle            

\begin{abstract}
Financial exchange operators cater to the needs
of their users while simultaneously ensuring compliance with the financial regulations.
In this work, we focus on the operators' commitment for
fair treatment of all competing participants.
We first discuss unbounded temporal fairness and
then investigate its implementation and infrastructure
requirements for exchanges.
We find that these requirements can be fully met only under
ideal conditions and argue that unbounded fairness in 
FIFO markets is unrealistic. 
To further support this claim, we analyse several real-world incidents
and show that subtle implementation inefficiencies and technical 
optimizations suffice to give unfair advantages to a minority of
the participants.
We finally introduce, $\epsilon$-fairness, a bounded definition of
temporal fairness and discuss how it can be combined with 
non-continuous market designs to provide equal participant treatment
with minimum divergence from the existing market operation.
\end{abstract}

\section{Introduction}
First-in-first-out (FIFO) markets process incoming messages in the
same temporal order they arrive to the matching engine. Thus,
reaction time to market events is one the most important
factors in the competition for scarce resources. 
Following the introduction of electronic trading, many firms sought to decrease
their reaction time to market events by automating their strategies (i.e., algorithmic 
trading) and by improving their technology (e.g., optimize their trading
models, establish faster network links, upgrade their software and hardware stacks). 
Over time, this evolved into a technological arms race between several firms 
competing for better placement at the order book\footnote{Orders who occupy the top of the 
order book have better access to liquidity~\cite{cont2014price}.}. 
The average reaction time has drastically reduced, initially from
seconds to milliseconds, and soon after down to microseconds and nanoseconds.

For any FIFO market race to be fair, participants must be treated equally and 
compete only on the basis of their respective reaction times. 
In this work, we argue that unconditionally equal treatment of the participants 
in FIFO markets is becoming increasingly difficult to achieve.
This is primarily due to FIFO's implementation and infrastructural requirements 
that are hard to consistently meet in practice. For example, exchanges
(whose participants may have nanosecond-scale reaction times) must ensure that their
infrastructure does not introduce any delays that could alter the 
relative arrival times of competing orders~\cite{deutschboerse2019,meltonFloor,melton2018fairware}.
However, networking equipment (e.g., network switches) is rarely certified to guarantee 
equal latency on all ports at a nanosecond scale~\cite{deutschboerse2019}.
To better understand the deployment complexities of FIFO order-matching,
we study several real-life incidents where exchanges failed to maintain
a level playing field for all market participants.
Most of these incidents were due to unintentional, subtle infrastructural inefficiencies 
but, in certain cases, participants intentionally further exacerbated
the timing discrepancies through technical\footnote{In this work, we use the term ``technical'' to refer to knowledge, machines, or methods used in science and industry. This is not to be confused with ``technical analysis'' that refers to the analysis of statistical trends.}
optimizations. 
Based on the insights from our survey, we then argue that unbounded 
temporal fairness is an elusive goal and introduce a bounded extension.
This more practical version of fairness could improve the microstructural 
transparency of financial exchanges as well as allow participants to determine if the 
degree of temporal fairness provided by an exchange is acceptable given their strategies.\\

\noindent Overall, this paper makes the following contributions:

\begin{itemize}
	\item Defines the implementation and infrastructure requirements that must be fulfilled by every exchange using an unbounded FIFO order-matching policy.
	\item Surveys several ``unfairness'' incidents from financial exchanges and investigates their root causes.
	\item Introduces a bounded version of fairness that improves the market transparency towards participants and could serve as a reference point for regulators.
\end{itemize}

\section{Preliminaries}\label{sec:preliminaries}
In this section, we introduce some fundamental concepts of electronic markets and outline
the operation of modern exchanges.

\subsection{Market Structure}\label{subsec:structure}
\begin{figure}[h]
    \centering
    \includegraphics[width=\columnwidth]{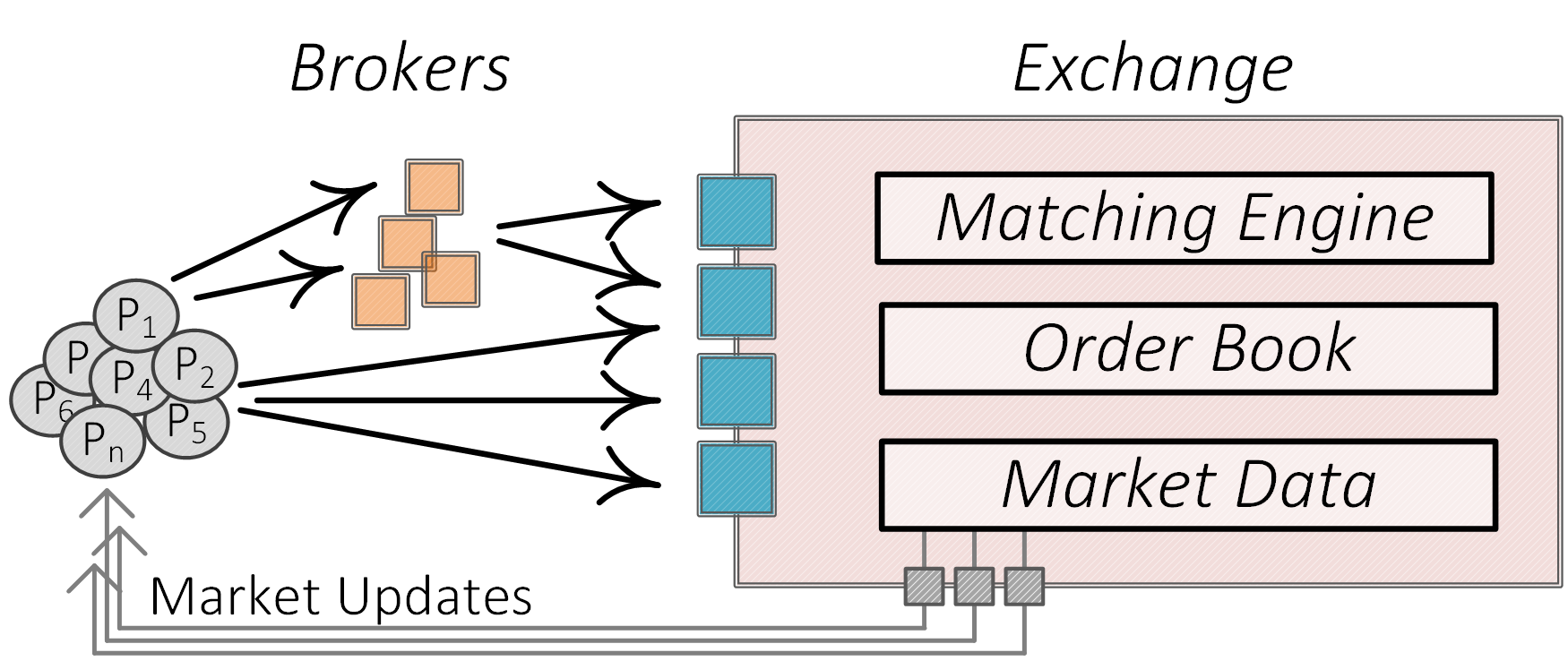}
    \caption{Illustration of the components of an electronic financial exchange.
    Participants submit their orders through their brokers (orange rectangles) or directly to the exchange's gateways (blue rectangles). The gateways then forward all valid orders to the matching engine, which pairs bids and offers and updates the order book with the remaining quantities. Market participants remain up to date with the market by subscribing to one of the available market feeds.}
    \label{fig:microstructure}
\end{figure}

Figure~\ref{fig:microstructure} illustrates the different actors and components of an electronic market and their interactions.
Market participants ($P_i$) submit their \textit{orders} to the exchange either through brokers or directly to the exchange's gateways. Brokers are used
by participants who trade low volumes, while those with higher volumes usually prefer to 
place their computers in the exchange's premises (i.e., colocation) and connect 
directly to the exchange (i.e., direct market access). In both cases, the exchange receives the incoming 
orders through its gateways which filter out invalid orders and
forward the rest to be matched. To balance the load and reduce latency
under heavy traffic conditions, exchanges tend to deploy several 
order gateways very close to the matching engine~\cite{meltonADSN}. 

Once an order is received, it is placed in the \textit{order book}
where the matching algorithm ranks, pairs and fills it with those of other participants.
\footnote{A more thorough treatment of the limit order book is provided by Gould et al.~\cite{gould2013}.}
Once a match is found, the order is closed and removed from the book, if the entire open quantity of the order has been filled. 
Otherwise, the book's record is updated to include only the remaining (unfilled) quantity. 
Participants remain up to date on their pending orders and the state of the order book
through the market's data feeds which are routed through the exchange's update servers. Depending on their feed subscription type, participants may receive only periodic snapshots of the order book or
near-realtime updates on the trading activity.

\subsection{Order-Matching Policies}\label{subsec:competition}
Order-matching policies specify the manner by which a financial exchange must process (e.g., rank, match, hide) the order messages
it receives from market participants. Currently, one of the most commonly used policies is FIFO (also known as first-come-first-serve matching)~\cite{Preis2011}. Under this policy, the matching engine processes order messages in the same 
temporal order in which the messages were received. Consequently,  in markets with scarce resources, participants who react fast to market events have higher allocation chances compared to participants with slower reaction times.

In modern exchanges, this speed-related competition has three different manifestations depending on the roles of the competing parties i.e., maker vs. maker, taker vs. taker and maker vs. taker\footnote{In finance literature, ``market makers'' provide liquidity to the market (i.e., post resting orders that populate the limit order book), while ``takers'' remove liquidity by placing orders that are `marketable' (so immediately match with makers' orders  causing those maker orders to be removed from the book). These roles are not fixed and participants transition from maker to taker depending on the side of the trade they take.}. 
In the first case, makers compete with other makers for the best position in the queue at each price level in the limit order book. An order that is earlier in the queue at a price level is much more likely to get filled than one positioned later in the queue~\cite{farmer2012,moallemi2014value}. Similarly, takers compete with other takers for favorably-priced bids and offers of scarce resources~\cite{farmer2012}. The first taker order to the market has access to the full offered volume, while subsequent ones have a lower chance of getting filled, as the volume decreases.
A maker who is trying to cancel their ``stale'' bid or offer may also compete with taker(s) who are trying to lift that bid or offer. This practice is called ``sniping'' and involves fast takers buying assets from or selling assets to slower makers at `stale' bid and offer prices~\cite{budish2015,farmer2012,kirilenko2017flash,mavroudis2019libra}. A more thorough analysis of matching policies can be found in~\cite{janecek2007matching}.

\section{Unbounded Temporal Fairness}\label{subsec:fairness}
In economic theory, fair races for a resource in contention are won by the ``fastest'' participant 
if the order book operates in a {\em continuous} FIFO manner~\cite{budish2015,harris2013}. 
Two market participants $P_A$ and $P_B$ are said to {\em compete} 
for a trading opportunity $op$ if responding to the same  
economic stimulus (e.g., market event), they both submit order messages that seek to capture $op$~\cite{angel2013fairness,mavroudis2019libra,melton2017fairness}.
We define as {\em reaction time} the time that elapses between 
the receipt of the stimulus by a participant $P$ and the submission of an order responding to 
the stimulus by $P$. 
In the rest of this paper, we adopt the position of~\cite{budish2015,harris2013} and assume that a
market participant's speed is a function of their (direct or indirect) investment in technology
i.e., the more a participant spends the faster they are expected to be. 
With this in mind, we can now provide a definition of temporal fairness in financial exchanges~\cite{mavroudis2019libra,melton2017fairness}:

\begin{definition}{Temporally Fair Race:}\label{def:fairrace}
Given a market event occurring at time $t_e$, a race between two 
participants $P_A$ and $P_B$ with reaction times $r_A$ and $r_B$
is fair, if for the arrival times of their orders ($t_A$ and $t_B$)
it holds that:
$t_A=t_e+r_A+l$ and $t_B=t_e+r_B+l$, where $l$ is constant.
\end{definition}

The constant $l$ represents the various transmission delays of the exchange's infrastructure
(e.g., network link latency for market updates, order transmission latency, order processing latency).
A more relaxed version of this definition that seeks to bring slower and faster participants
onto ``equal footing'' has been discussed in~\cite{mavroudis2019libra,melton2017fairness,meltonparadoxical,melton2018fairness}.
We can now define when an exchange is temporally fair.

\begin{definition}{Temporally Fair Exchange:}\label{def:fairness}
A financial exchange is considered temporally fair, if all the races between 
all possible pairs of its participants are always \textit{temporally fair}.
\end{definition}

\section{FIFO Requirements}\label{sec:reqs}
Intuitively, FIFO matching appears to conform with the above definition of temporal fairness by-design
(see Section~\ref{subsec:fairness}) as orders are ranked and matched based on their relative 
arrival time at the exchange's engine. However, a FIFO market design can maintain its fairness
properties only if orders always reach the matching engine in the same relative order 
they were sent. Similarly, market updates should reach all the colocated participants
simultaneously so as to ensure that no one has more time to react. More formally,
a FIFO deployment is temporally fair if it remains compliant with the following three
requirements at all times:\\

\noindent\textit{Consistent \& Simultaneous market updates}\\
Given a market event $e$ occurring at time $t_e$ and two colocated participants
who receive the corresponding market update at times $t_A$ and
$t_B$ respectively, it must always hold that $t_A - t_e$ = $t_B - t_e$,
for every possible pair of participants.\\

\noindent\textit{Preserve relative submission order}\\
Given two colocated participants whose orders were sent at times
$t_A$ and $t_B$ and arrived at the matching engine at times $t_A'$
and $t_B'$, it must always hold that $t_A'-t_A = t_B'-t_B$.\\

\noindent\textit{Honor price-time priority}\\
Given an order $o_A$ that arrives at the matching engine at time $t_A$
and a competing order\footnote{The term ``competing orders'' refers to lit orders 
for the same instrument that reside on the same order book side and price level.}
$o_B$ that arrives at $t_B$, if $t_A<t_B$, it must always hold that $o_A$ will
be executed before $o_B$ (unless $o_A$ is explicitly cancelled by the participant).

\section{Practical Considerations}\label{sec:incidents}
As outlined in Sections~\ref{subsec:fairness} and \ref{sec:reqs}, temporally 
fair markets require that participants compete solely on the basis of their reaction time while 
every other factor of the system remains constant for all of them.
However, the practicality of these strict requirements is questionable.
We now survey various incidents from the relevant literature as well as
legal reports and show that in various cases participants experienced
favourable or unfavourable order submission latency or market update delay 
due to technical inefficiencies.\\

\noindent\textbf{Infrastructure Jitter.}
Electronic exchanges comprise a plurality of hardware and software components that
exhibit small non-constant variations in their processing times. For example, modern 
computers are optimized for maximum performance (i.e., instruction throughput) and thus do
not guarantee deterministic execution times. This is due to speculative execution, 
cache eviction in the presence of competing processes, charge-to-read and data prefetching
that contribute to variances in the process execution runtimes in a range between
a few nanoseconds to a few hundred microseconds~\cite{Metamako,Godbolt18}.
Furthermore, distributed network architectures (such as those used by major financial exchanges)
also introduce discrepancies in the processing times and asymmetries in the data
dissemination speed. This is attributed to the various performance differences between  
replicated components. For instance, an order gateway may be significantly less crowded 
compared to the rest of the gateways, thus providing favourable submission times
to its users~\cite{SEBI2109}.
To address this, market operators strive to maintain their systems perfectly symmetrical and
load balanced at all times with nanosecond~\cite{deutschboerse2019} precision.
While this can reduce the magnitude of the problem, differences cannot be fully 
eliminated even if the exchange uses and fine-tunes the exact same hardware and software in all of its
replicated components.\\

\noindent\textbf{Uneven Information Dissemination.}
Another potential point of failure with regards to temporal fairness
are the market operations that involve information dissemination.
Such operations, if not implemented carefully, have the potential to 
introduce discrepancies in the times that information is received by
different participants. For example, Tick-by-Tick data feeds
update participants after every change in the order book and 
must be received by all subscribed participants simultaneously.
However, perfect synchronicity is technically challenging to achieve in practice.
In a relevant incident, the data feed servers at the National Stock Exchange 
(NSE) of India were reportedly transmitting the updates in a sequential, non-randomized manner
to the subscribed participants, while the order gateways differed in capabilities
and loads~\cite{SEBI2109}. In particular, market updates were transmitted in 
the same order participants logged in (on a per server basis), thus
enabling participants with knowledge of this technicality to gain an advantage by logging in early.
This resulted in unfair races as some participants were consistently receiving market updates
before everyone else (Requirement 1 in Section~\ref{sec:reqs}).
From a technical perspective, one-shot multicast could potentially decrease
the latency discrepancies by transmitting all the updates \textit{almost} simultaneously\footnote{Multicast based on fanout-spitting or application-layer multicast overlay services suffer from delay problems similar to those of unicast~\cite{prabhakar1997multicast}.}.
Unfortunately, even in this case, the inter-arrival discrepancies are reduced but not eliminated~\cite{klocking2001reducing}.
During the same incident, participants were also found to actively delay updates to others by unnecessarily occupying positions in the queue of update servers they were not using~\cite{SEBI2109}. Similar practices have been reported in various other exchanges and markets~\cite{ebsunfair2012,ebsnewrules,ebspatent2007}.\\

\noindent\textbf{Request Broadcasting.}
As discussed in Section~\ref{subsec:structure}, major exchanges have a modular (cf. monolithic) architecture 
with several replicated components to improve their responsiveness and minimize their downtime. However, component replication
can have an impact on fairness as there may be fluctuations in the performance of the different components over time.
Participants can take advantage of such discrepancies even if they have no information about the performance of the individual
components (e.g., order gateways). In past incidents, participants were found to simultaneously broadcast copies
of their orders to all available gateways~\cite{CME14}.
This practice improves the likelihood that their order will be routed through one of the faster (e.g., less loaded) gateways 
thus outracing competing participants who do not replicate their orders (Requirement 2 in Section~\ref{sec:reqs}).
Various exchanges have taken measures to counter order replication, usually by limiting the number of simultaneous connections that are 
permitted per participant~\cite{CME18,CME19,ebsunfair2012,ebsnewrules}.\\

\begin{figure}[h]
    \centering
    \includegraphics[width=\columnwidth]{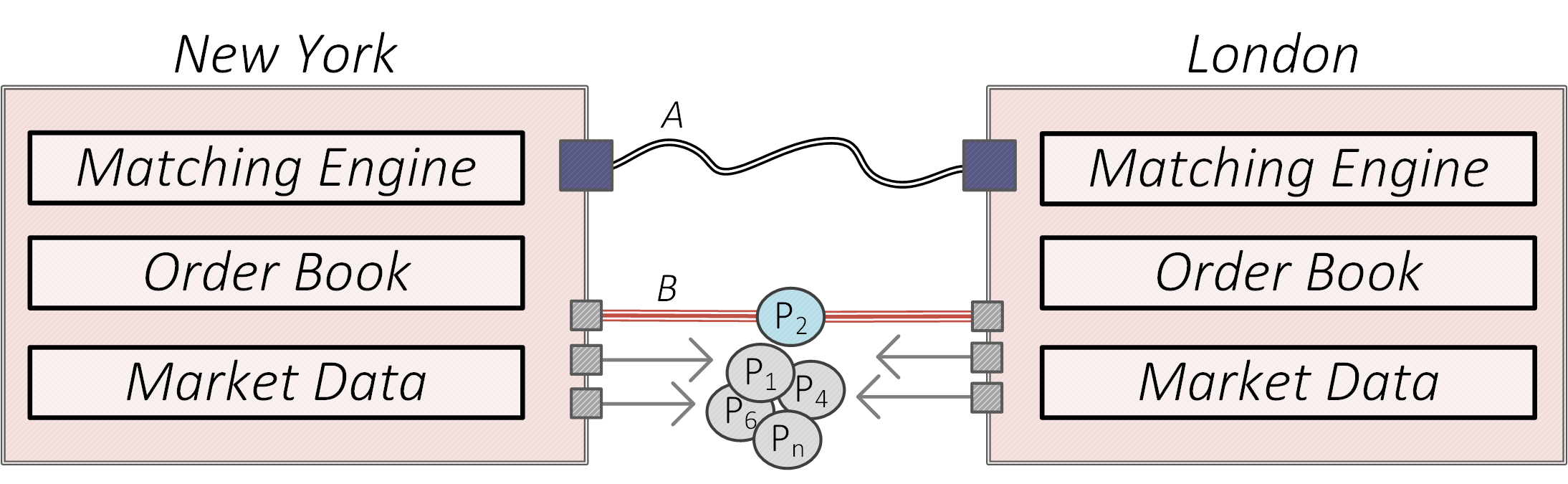}
    \caption{Illustration of how out-of-band channels can impair fairness in exchanges with decentralized
    order books. In such exchanges, orders submitted at the local order book (e.g., New York) may be routed to one of the geographically distributed nodes (e.g., London) that offers the best price.
    The nodes route orders to each other through a direct high-bandwidth
    network link ($A$), while participants are updated through the market data feeds of each order book. 
    However, this structure assumes that there is no link $B$ that provides lower latency than $A$. If such a link becomes available, participants can use it to front-run orders as they are routed across the nodes.}
    \label{fig:distr}
\end{figure}

\noindent\textbf{Out-of-band Channels.}
A common practice for market participants is to use private
fast network links to aggregate market data from multiple venues so as to take better
informed trading decisions. In centralized exchanges, this 
practice does not violate any of the fairness requirements as the order submission
latency and the update-feed delays remain constant.
However, fast external links may have an adverse effect on the fairness of exchanges with decentralized order books.
Such exchanges maintain several geographically-distributed nodes (i.e., order books) 
that use a network link to route incoming orders to the node that currently offers
the best price (Figure~\ref{fig:distr}). Thus, an incoming order that gets submitted
in London may be routed to New York to be matched. However, this assumes that any
participant who observes an incoming order in London will not be able to 
reach the New York book before the order is routed there.
While in most cases this is a reasonable assumption (such very low-latency links across continents are limited), the ICAP EBS venue's~\cite{ebspatent2007,ebsbypass2015} inter-region routing link was outperformed by the Hibernia Express link, thus advantaging participants who had subscribed to it. To address the problem,  the operator introduced a speedbump (i.e., a short constant delay) that alleviated the latency advantage provided by the faster network link~\cite{ebsbypass2015}.\\

\noindent\textbf{Communication Protocol Violations.}
In networks that use store-and-forward switching, each complete transmission unit (e.g., frame, packet)
is copied to the switch's memory buffer and checked for 
potential errors (e.g., cyclic redundancy checks) before it gets forwarded
to the next node. Therefore,
the processing time of each message by the switch depends on its 
size as the whole unit must be received before any
checks are performed. In 2015, a market participant was found to  
exploit this process by intentionally truncating data fields in their
outgoing orders. This resulted in invalid data in some non-critical 
fields but gave them an advantage over other competing participants as 
switches would process their orders faster~\cite{CME14,CME17}.

Participants have been also found to take advantage of 
the inherent properties of communication protocols. In networks, 
large messages are often split into fixed-size fragments that are sent to their 
destination sequentially. The recipient stores the incoming
fragments into a memory buffer and combines them to reconstruct the original message. 
However, message fragmentation may result in ambiguity with regards to the 
arrival times of incoming messages. This is of particular importance
in electronic exchanges as fragments of two competing orders (i.e., messages) may arrive interweaved. 
To resolve this, matching engines have introduced tie-breaking policies that allow them 
to determine the relative order of incoming orders in a consistent manner.
One such policy timestamps incoming orders based on the arrival time
of their first fragment (e.g., IP packet). This timestamp is then used to 
determine their relative precedence in the order book.
Unfortunately, this policy is prone to \textit{optimistic messaging},
a form of technical gaming. Optimistic messaging relies on the
fact that orders are timestamped with the arrival time
of the first fragment but cannot be reconstructed
until all fragments have arrived~\cite{CMEOptimistic,Hurd,deutschboerse2019}.
A strategic participant can exploit this to establish precedence in the
book for orders they may intend to submit in the future.
For example, given an economic event that is to occur at a specific time $t_e$, 
a participant preemptively sends a fragment of an order shortly before $t_e$
(a few milliseconds early) to establish precedence in the order book. Then, 
depending on the outcome of the event, the participant can decide to 
trade on the news and transmit the rest of the order's fragments, or drop the
incomplete order (e.g., by invalidating the network/application-layer checksum in the remaining fragments).
Depending on the variant used by the participant, the pre-event fragments may
contain the fields of the order that are not trade-specific (e.g., participant ID)
or may contain event-depend data if the event has only a few potential outcomes.\\

\noindent\textbf{Unintended Order Interactions.}
Electronic markets paved the way for sophisticated order types that
realize conditional interactions with the order book. However, the 
additional complexity introduced by those orders, made it more difficult 
for operators to account for all possible interaction scenarios.
For example, certain order types (e.g., intermarket sweep orders, hide and light orders) could under specific conditions
overtake other orders who preceded them in the execution queue (Requirement 3 in Section~\ref{sec:reqs})~\cite{bodek2013problem,mavroudis2019market}.
While these order types were available to everyone, their existence and operation 
was poorly communicated thus resulting in an unfair asymmetry between the participants who
were aware of their operation and those who did not~\cite{ubsSEC,Buti13sub-pennyand,bodek2013problem,australiasic}.
In another occasion, the interaction between two order-types on New York Stock Exchange
enabled participants to use pegging orders to detect the presence (but not the volume) of
hidden orders~\cite{SECNYSE2018}. As a result, participants who exploited this design flaw could  
retrieve additional information on the current state of the order book compared to those that
relied only on the market feeds. While this order interaction was a bug, it highlights 
how unintended corner-cases may occur in exchanges with complex microstructure.

\section{Bounded temporal fairness}\label{sec:efairness}
As discussed in Section~\ref{sec:reqs}, an exchange that guarantees fairness unconditionally must ensure that 
all updates arrive simultaneously to all its colocated participants.
However, unlike \textit{price}, \textit{time} is continuous and thus simultaneity 
is unrealistic, especially considering the various hardware and software imperfections. 
We thus argue that \textit{unbounded} fairness is an elusive goal.
Instead, fairness should be considered with regards to a clearly defined \textit{reference frame}. 
We now introduce bounded versions of the definitions given in Section~\ref{subsec:fairness} that
provide such a reference frame.

\begin{definition}{$\epsilon$-Bounded Temporally Fair Race:}\label{def:fairrace}
Given a market event occurring at time $t_e$, a race between two 
participants $P_A$ and $P_B$ with reaction times $r_A$ and $r_B$
is $\epsilon$-fair, if for the arrival times of their orders ($t_A$ and $t_B$)
it holds that:
$t_A=t_e+r_A+l \pm \epsilon/2$ and $t_B=t_e+r_B+l\pm \epsilon/2$, where $l$ is constant.
\end{definition}

As before, the constant $l$ represents the various transmission delays of the exchange's infrastructure,
while $\epsilon$ quantifies the time scale at which the current infrastructure of the exchange can provide
fair treatment between the participants. For example, a switch that consistently offsets one of its ports by
1ms can provide $\epsilon$-fairness for $\epsilon=1ms$, as the faster participant in races
where $\|r_A - r_B \| > 1ms$ will still win the race (assuming the rest of the infrastructure exhibits no jitter).
For brevity, we refer to $\epsilon$-bounded temporally fair races as $\epsilon$-fair.

\begin{definition}{$(\epsilon,\delta)$-Bounded Temporally Fair Exchange:}\label{def:fairness}
A financial exchange is considered temporally fair with respect to bounds $\epsilon$ and $\delta$,
if the races between all possible pairs of its participants are $\epsilon$-fair
with probability greater than or equal to $\delta$.
\end{definition}

In the above definition, $\delta$ represents the probability
that a race between two orders will be $\epsilon$-fair.
We introduced $\delta$, as past works have shown that the 
probability distributions of the latency in distributed systems
(both in financial exchanges and in other applications)
exhibit long tails~\cite{li2014tales,deutschboerse2019}.
In a practical setting, the operator can derive $\epsilon$
and $\delta$ by monitoring the discrepancies 
occurring in their system. Some techniques and measurement
tools for this purpose have been discussed in~\cite{deutschboerse2019}.

Overall, the above definitions provide a more practical notion of temporal fairness
as they relax the requirements for strict simultaneity in market updates and
complete precedence-preservation in orders. \textit{Bounded} fairness can 
improve the transparency of financial exchanges and allow
participants to determine if the degree of temporal fairness provided
by an exchange is acceptable given their strategies. Moreover, $\epsilon$ and $\delta$ could 
serve as a well-defined reference point for regulators and 
could drive a positive competition between exchanges.

Note, however, that these definitions do not make a statement about the 
races that take place beneath the $\epsilon$ boundary. This leaves a blind spot, 
where all the races with reaction-time differences less than $\epsilon$ could 
be consistently won by the slowest competing participant.
Operators can try to decrease $\epsilon$ (and increase $\delta$)
by upgrading their infrastructure but similarly participants
are also getting consistently faster and more sensitive to subtle differences.
We argue that non-continuous market designs at $\epsilon$ timescales could solve
this problem by uniformly distributing the
victories between participants for races beneath $\epsilon$~\cite{mavroudis2019libra,melton2017fairness,meltonparadoxical,melton2018fairness}.
In these market designs, the majority of the participants will experience a FIFO market, while the 
small minority of ultra-fast traders will be still guaranteed at least equal treatment at all scales.

\section{Conclusions \& Future Directions}\label{sec:conclusion}
This work discusses the gap between economic theory
and the unavoidable complexities that emerge when 
theoretically ``fair'' market models are implemented. 
We focus on FIFO order-matching and its
infrastructural and implementation requirements
that ``fair'' deployments must fulfill.
Our survey of the literature and relevant sources
suggests that technical challenges make full 
compliance with these strict requirements unrealistic.
Instead, we propose a bounded version of fairness 
that takes into account the unpredictability of hardware
and software systems. The bounds clearly define the 
degree of fairness that participants can expect from an
exchange and can be tightened further as the operators 
refine their infrastructure. For cases that fall outside
these bounds (i.e., very low latency competition), 
discrete-time policies can be applied to guarantee at
least equal participant treatment. 

\subsubsection*{Acknowledgments}
We thank the attendees of the workshop for the insightful discussion and the valuable feedback.
Moreover, we would like to thank David Kohan Marzagão for his input with regards to our definitions of fairness. Vasilios Mavroudis was partially supported by University College London and BinanceX through the Tradescope project.

\bibliographystyle{splncs04}
\bibliography{bibliography}

\end{document}